\documentclass{svproc}
\usepackage{amsmath,bm,bbm,latexsym,soul,nicefrac,young,youngtab}
\usepackage{graphicx,subfigure,multirow,dcolumn}

\usepackage{url}

\newcommand{\fref}[1]{Fig.~\ref{#1}}

\newcommand{\tref}[1]{Tab.~\ref{#1}}

\usepackage{array}
\usepackage{stackengine}
\newcommand\xrowht[2][0]{\addstackgap[0.5\dimexpr#2\relax]{\vphantom{#1}}}

\begin{document}
\mainmatter              
\title{Joint gravitational wave detection by TianQin and LISA}
\titlerunning{Joint detection by TianQin and LISA}  
%
\author{Alejandro Torres-Orjuela\inst{1}}
\authorrunning{Alejandro Torres-Orjuela} 
\institute{Department of Physics, The University of Hong Kong, Pokfulam Road, Hong Kong,\\
\email{atorreso@hku.hk}}

\maketitle              

\begin{abstract}
We study the detection accuracy of double white dwarfs (DWDs), stellar-mass black hole binaries (SBHBs), light and heavy intermediate mass ratio inspirals (IMRIs), extreme mass ratio inspirals (EMRIs), massive black hole binaries (MBHBs), and the stochastic gravitational wave background (SGWB) of astronomical origin for TianQin, LISA, and joint detection. We use a Fisher matrix analysis and consider for each source the averaged detection accuracy over a realistic range of parameters. We find that on average TianQin obtains more accurate parameter estimation for DWDs and light IMRIs, LISA for heavy IMRIs, EMRIs, MBHBs, and the galactic foreground of the SGWB, and both contribute similarly to the detection of SBHBs and the extra-galactic SGWB. Nevertheless, for all sources joint detection allows setting tighter constraints on most parameters highlighting its importance for future detection.

\end{abstract}

\section{Introduction}

Gravitational wave (GW) detection has opened a new window to study compact objects~\cite{GWTC1,GWTC2,GWTC3} although it is currently restricted to the high-frequency band above $1\,{\rm Hz}$ covered by ground-based laser interferometry detectors LIGO, Virgo, and KAGRA~\cite{ligo_2015,virgo_2012,kagra_2019} and the $\rm nHz$ band accessible through pulsar timing arrays~\cite{ipta_2013,nanograv_2013,ppta_2013,cpta_2016,epta_2016,inpta_2018}. The detectable spectrum will be expanded in the mid-2030s to the low-frequency ($\rm mHz$) band by space-based laser interferometry detectors TianQin, LISA, and Taiji~\cite{tq_2016,lisa_2017,taiji_2015,tq_2024,lisa_2022a} while the intermediate band ($\rm dHz$) will be opened by space-based laser interferometry detectors like DECIGO~\cite{decigo_2021}, ground-based and space-based atom interferometry detectors like AION, ZAIGA, and AEDGE~\cite{aion_2020,zaiga_2020,aedge_2020} as well as lunar GW detectors like LGWA~\cite{ajith_amaro-seoane_2024} in coming years.

TianQin and LISA are two space-based laser interferometer detectors with a similar design but differ in key aspects. Due to the differences in their design, TianQin and LISA can detect the same sources but obtain different accuracies. These sources include double white dwarfs (DWDs) which are the most abundant class of compact stellar mass binaries in our galaxy~\cite{nelemans_yungelson_2001a,yu_jeffery_2010,breivik_coughlin_2020}, stellar-mass black hole binaries (SBHBs) formed through co-evolving massive stars~\cite{vanbeveren_2009,belczynski_dominik_2010,kruckow_tauris_2018} or dynamical assembling in dense stellar systems~\cite{portegies-zwart_mcmillan_2000,gultekin_miller_2004,tagawa_haiman_2020}, light intermediate mass ratio inspirals (IMRIs) formed by a stellar-mass black hole (BH) orbiting an intermediate-mass BH (IMBH)~\cite{will_2004,amaro-seoane_gair_2007,arca-sedda_amaro-seoane_2021}, heavy IMRIs formed by an IMBH orbiting a massive BH~\cite{basu_chakrabarti_2008,arca-sedda_gualandris_2018,derdzinski_dorazio_2021}, extreme mass ratio inspirals (EMRIs) formed by a small compact object inspiraling into a MBH~\cite{amaro-seoane_2018a,mapelli_ripamonti_2012}, MBH binaries (MBHBs) residing in the center of galaxies~\cite{komossa_2003,milosavljevic_merritt_2003a,milosavljevic_merritt_2003b}, and the stochastic GW background (SGWB) resulting from the combination of a multitude of unresolved binaries~\cite{allen_romano_1999,romano_cornish_2017,christensen_2019}.

We study the detection of astrophysical GW sources by TianQin, LISA, and joint detection using a Fisher matrix analysis~\cite{coe_2009,torres-orjuela_huang_2023,wang_han_2021}. The inverse of the Fisher Matrix $C = F^{-1}$ is a linear estimate of the measurement errors that asymptotes the true error in the limit of high signal-to-noise ratio. The detection accuracy of the joint detection is obtained by combining the results of the two independent detectors $F_{\rm Joint} = F_{\rm TianQin} + F_{\rm LISA}$ and then inverting $F_{\rm Joint}$. We focus on comparing the averaged detection accuracy of TianQin, LISA, and joint detection. However, the detection accuracy of a parameter can vary significantly depending on its specific value. Therefore, the results shown here shall only give an idea of the detection accuracy for the different detection scenarios and we refer the reader to Ref.~\cite{torres-orjuela_huang_2023} for more details. Moreover, when calculating the errors over the parameter range we only vary the considered parameter and fix the other parameters to fiducial values. The distributions used for the parameters are either uniform in linear space (lin), uniform in log-space (log), or uniform in cosine (cos).

\section{Detection by TianQin, LISA, and joint detection}

\subsection{Double white dwarfs}\label{sec:dwd}

\begin{figure*}
    \centering
    \begin{minipage}{0.48\textwidth}
        \centering
        \includegraphics[width=0.95\textwidth]{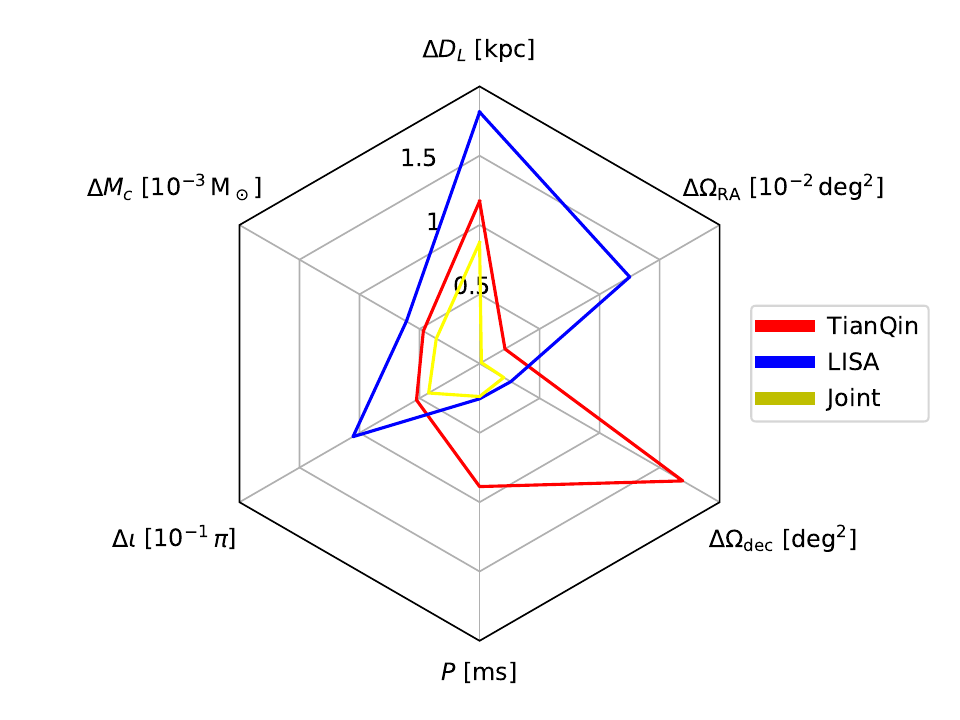}
    \end{minipage}\hfill
    \begin{minipage}{0.48\textwidth}
        \centering
        \includegraphics[width=0.95\textwidth]{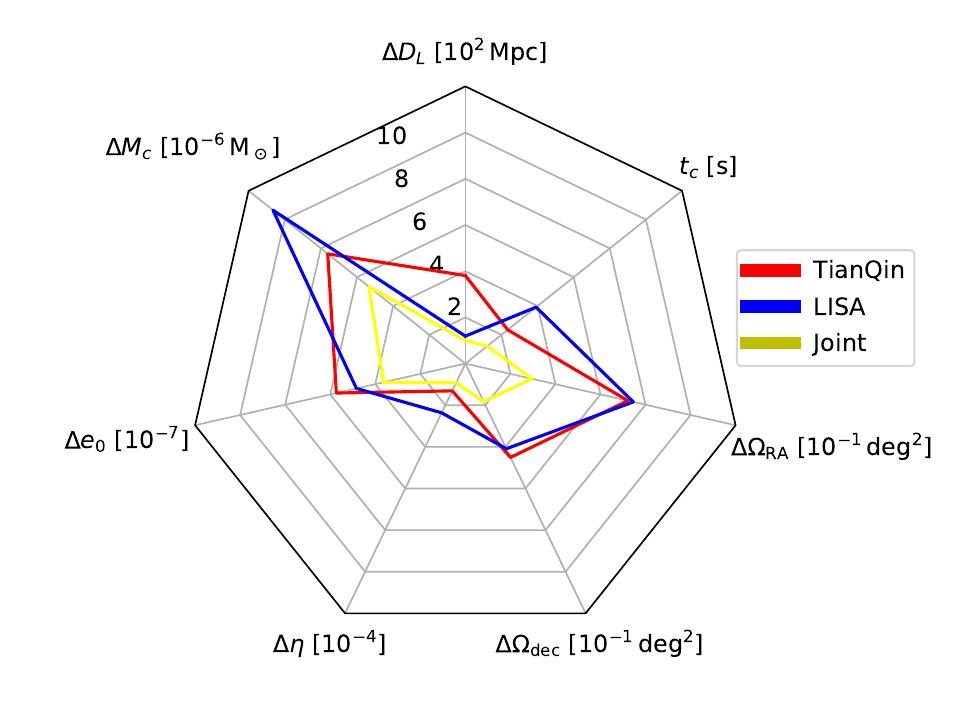}
    \end{minipage}
    \caption{
        The averaged absolute error for the parameters of DWDs (left) and SBHBs (right) measured by TianQin (red), LISA (blue), and joint detection (yellow).
    }\label{fig:dwdsbhb}
\end{figure*}

The parameters we consider for DWDs are the luminosity distance $D_L$, the chirp mass $M_c$, the inclination of the orbit relative to the line-of-sight $\iota$, the orbital period of the binary $P$, and the sky localization $\Omega$. We consider the sky localization error as a function of the declination $\rm dec$ and the right ascension $\rm RA$ where one is fixed while the other is varied. The fiducial values, the parameter ranges considered and the distributions used, as well as the average errors are shown in \tref{tab:dwd}. Note that the average errors for $\rm dec$ and $\rm RA$ refer to the sky localization error $\Delta\Omega$ as a function of these two parameters and are given in units of $\rm deg^2$. We see on the left-hand side of \fref{fig:dwdsbhb} that TianQin constrains $D_L$, $M_c$, $\iota$, and $\Omega_{\rm RA}$ better than LISA while LISA can set tighter constraints than TianQin for $P$ and $\Omega_{\rm dec}$. Therefore, we find that although the detection of single parameters of a DWD might be dominated by one of the two detectors, joint detection yields significantly better results when studying all parameters combined. Note that as in real detection, all parameters are measured at the same time the improvements from joint detection are even more significant than they appear here. Last, we point out that the relatively big error in $\Omega_{\rm dec}$ compared to $\Omega_{\rm RA}$ for TianQin can be attributed to the particular choice of fiducial parameters and similar accuracies to those of $\Omega_{\rm RA}$ can be expected for most sources.

\begin{table*}\centering
    \caption{The fiducial values, the parameter ranges, and the averaged absolute errors in TianQin, LISA, and joint detection for DWDs.}\label{tab:dwd}
    \begin{tabular}{|m{1.6cm} || m{1.6cm} | m{1.6cm} | m{1.6cm} | m{1.6cm} | m{1.6cm} | m{1.6cm}|}
        \hline\xrowht[())]{7pt}
        & $D_L$ [kpc] & $M_c$ [$\rm M_\odot$] & $\iota$ [$\pi$] & $P$ [$\rm s$] & dec & RA \\ \hline\hline\xrowht[()]{7pt}
        Fiducial & 8.5 & 0.25 & 1/3 & 200 & $-5.6^\circ$ & $17.8^{\rm h}$ \\ 
        \hline\xrowht[()]{7pt}
        Range & 0.01-50 & 0.1-1.25 & 0-0.5 & 10-500 & $-90^\circ$-$+90^\circ$ & $0^{\rm h}$-$24^{\rm h}$ \\
        \hline\xrowht[()]{7pt}
        Distr. & log & lin & cos & lin & lin & lin \\
        \hline\xrowht[()]{7pt}
        TianQin & 1.2 & $4.6\times10^{-4}$ & $5.3\times10^{-2}$ & $8.9\times10^{-4}$ & 1.7 & $2.1\times10^{-3}$ \\
        \hline\xrowht[()]{7pt}
        LISA & 1.8 & $6.1\times10^{-4}$ & $1.1\times10^{-1}$ & $2.5\times10^{-4}$ & $2.6\times10^{-1}$ & $1.3\times10^{-2}$ \\
        \hline\xrowht[()]{7pt}
        Joint & $8.7\times10^{-1}$ & $3.6\times10^{-4}$ & $4.2\times10^{-2}$ & $2.4\times10^{-4}$ & $2.0\times10^{-1}$ & $1.5\times10^{-4}$ \\
        \hline
    \end{tabular}
\end{table*}

\subsection{Stellar-mass black hole binaries}\label{sec:sbhb}

For SBHBs, we consider $D_L$, $M_c$, the binary's eccentricity at $10\,{\rm mHz}$ $e_0$, the symmetric mass ratio $\eta$, $\Omega$, and time to coalescence $t_c$. The errors are calculated over a parameter range where we only vary the considered parameter and fix the other parameters to fiducial values. The fiducial values, the parameter range, and the distributions applied, as well as the averaged absolute errors in TianQin, LISA, and joint detection can be found in \tref{tab:sbhb}. From the right side of \fref{fig:dwdsbhb} we see that $M_c$, $\eta$, and $t_c$ are constrained better by TianQin while $D_L$ and $e_0$ have smaller errors in LISA. Moreover, we see that $\Omega_{\rm dec}$ and $\Omega_{\rm RA}$ are constrained to a similar level by both detectors, and the small difference can be mainly attributed to the particular choice of fiducial values. In general, all parameters are constrained to a similar level by TianQin and LISA leading to a significant improvement in joint detection - in particular, for $\Omega$, $e_0$, and $M_c$.

\begin{table*}\centering
    \caption{The fiducial values, the parameter ranges, and the averaged absolute errors in TianQin, LISA, and joint detection for SBHBs.}\label{tab:sbhb}
    \begin{tabular}{|m{1.45cm} || m{1.35cm} | m{1.35cm} | m{1.35cm} | m{1.45cm} | m{1.35cm} | m{1.35cm} | m{1.35cm}|}
        \hline\xrowht[())]{7pt}
        & $D_L$ [Mpc] & $M_c$ [$\rm M_\odot$] & $e_0$ & $\eta$ & dec & RA & $t_c$ \\ \hline\hline\xrowht[()]{7pt}
        Fiducial & 200 & 50 & 0.05 & 0.125 & $+30^\circ$ & $12^{\rm h}$ & $2.5\,{\rm yr}$ \\ 
        \hline\xrowht[()]{7pt}
        Range & 100-530 & 3.5-100 & 0.01-0.1 & 0.001-0.25 & $0^\circ$-$+90^\circ$ & $0^{\rm h}$-$24^{\rm h}$ & $1\,{\rm yr}$-$5\,{\rm yr}$  \\
        \hline\xrowht[()]{7pt}
        Distr. & log & log & lin & log & cos & lin & log \\
        \hline\xrowht[()]{7pt}
        TianQin & $380$ & $7.6\times10^{-6}$ & $5.7\times10^{-7}$ & $1.3\times10^{-4}$ & $4.5\times10^{-1}$ & $7.2\times10^{-1}$ & $2.3\,{\rm s}$  \\
        \hline\xrowht[()]{7pt}
        LISA & $110$ & $1.1\times10^{-5}$ & $4.8\times10^{-7}$ & $2.4\times10^{-4}$ & $4.1\times10^{-1}$ & $7.5\times10^{-1}$ & $3.9\,{\rm s}$  \\
        \hline\xrowht[()]{7pt}
        Joint & $100$ & $5.3\times10^{-6}$ & $3.6\times10^{-7}$ & $9.2\times10^{-5}$ & $1.8\times10^{-1}$ & $2.9\times10^{-1}$ & $1.2\,{\rm s}$  \\
        \hline
    \end{tabular}
\end{table*}

\subsection{Light intermediate mass ratio inspirals}\label{sec:limri}

\begin{figure*}
    \centering
    \begin{minipage}{0.48\textwidth}
        \centering
        \includegraphics[width=0.95\textwidth]{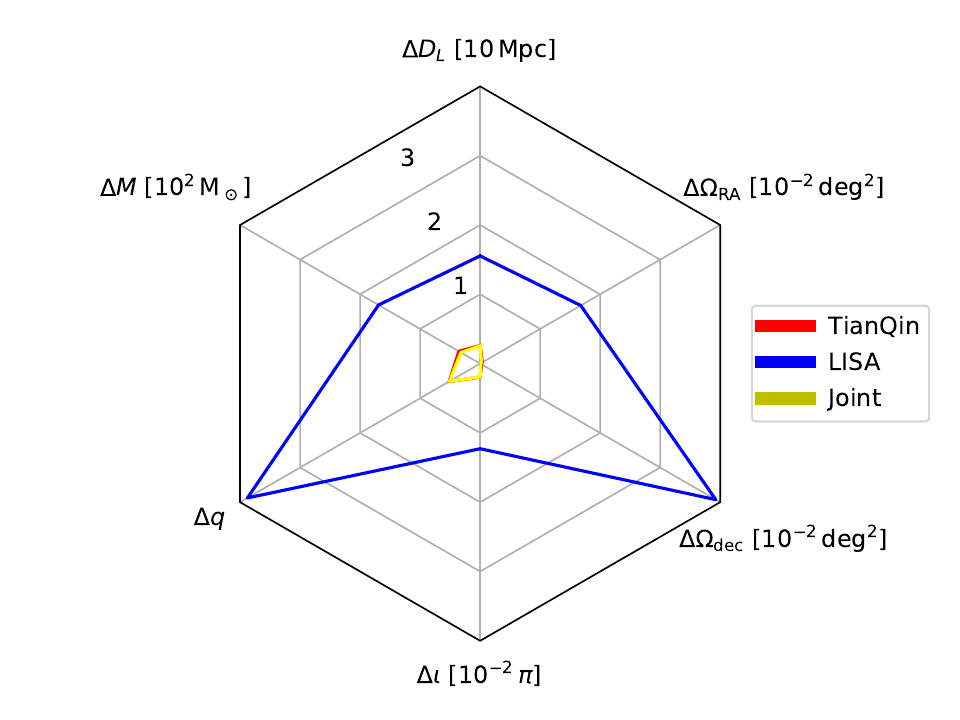}
    \end{minipage}\hfill
    \begin{minipage}{0.48\textwidth}
        \centering
        \includegraphics[width=0.95\textwidth]{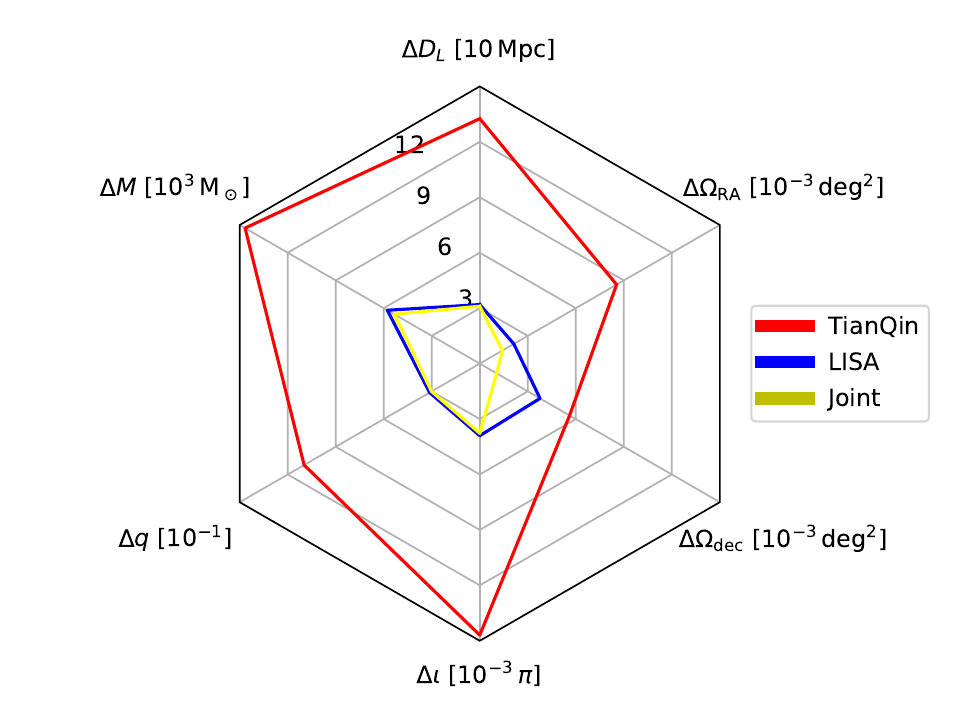}
    \end{minipage}
    \caption{
        The averaged absolute error for the parameters of light IMRIs (left) and heavy IMRIs (right) measured by TianQin (red), LISA (blue), and joint detection (yellow).
    }\label{fig:imris}
\end{figure*}

We consider $D_L$, the total mass $M$, the mass ratio $q$, $\iota$, and $\Omega$ as the parameters of light IMRIs, where the fiducial values, the parameter ranges used, and their distribution are shown in \tref{tab:limri}. The averaged absolute errors for TianQin, LISA, and joint detection can be found in the same table. The left-hand side of \fref{fig:imris} shows the averaged detection error for light IMRIs. We see that the detection accuracy of TianQin is significantly better than the detection accuracy of LISA; often differing by almost an order of magnitude. Therefore, the accuracy of joint detection is dominated by TianQin but LISA can contribute to an improved detection of $\Omega$ and $M$.

\begin{table*}\centering
    \caption{The fiducial values and the parameter ranges as well as the averaged detection errors in TianQin, LISA, and joint detection for light IMRIs.}\label{tab:limri}
    \begin{tabular}{|m{1.6cm} || m{1.6cm} | m{1.6cm} | m{1.6cm} | m{1.6cm} | m{1.6cm} | m{1.6cm}|}
        \hline\xrowht[())]{7pt}
        Parameter & $D_L$ [Mpc] & $M$ [$\rm M_\odot$] & $q$ & $\iota$ [$\pi$] & dec & RA \\ \hline\hline\xrowht[()]{7pt}
        Fiducial & 1 & 5000 & 100 & 1/3 & $0^\circ$ & $12^{\rm h}$ \\ 
        \hline\xrowht[()]{7pt}
        Range & 0.001-100 & 550-$10^5$ & 10-5000 & 0-0.5 & $0^\circ$-$+90^\circ$ & $0^{\rm h}$-$24^{\rm h}$ \\
        \hline\xrowht[()]{7pt}
        Distr. & log & log & log & lin & lin & lin \\
        \hline\xrowht[()]{7pt}
        TianQin & $2.6$ & $35$ & $5.1\times10^{-1}$ & $1.9\times10^{-3}$ & $2.4\times10^{-4}$ & $3.6\times10^{-4}$ \\
        \hline\xrowht[()]{7pt}
        LISA & $15.5$ & $169$ & $3.9$ & $1.2\times10^{-2}$ & $3.9\times10^{-2}$ & $1.7\times10^{-2}$ \\
        \hline\xrowht[()]{7pt}
        Joint & $2.6$ & $32$ & $5.1\times10^{-1}$ & $1.9\times10^{-3}$ & $1.3\times10^{-4}$ & $2.4\times10^{-4}$ \\
        \hline
    \end{tabular}
\end{table*}

\subsection{Heavy intermediate mass ratio inspirals}\label{sec:himri}

For heavy IMRIs, we consider the same parameters as for a light IMRI, namely $D_L$, $M$, $q$, $\iota$, and $\Omega$ and again compute the errors over a parameter range varying only the considered parameter while fixing the other parameters to fiducial values. In \tref{tab:himri} we show the fiducial values, the range of the parameters, the distributions used for them, and the averaged detection error for TianQin, LISA, and joint detection. We, further, see on the right side of \fref{fig:imris} that the detection accuracy in LISA is significantly better than TianQin's although the difference is less than one order of magnitude. Therefore, the constraints set by the joint detection are similar to those of LISA for $D_L$, $M$, $q$, and $\iota$. For $\Omega$ the detection accuracy of the joint detection is also dominated by LISA but TianQin can considerably contribute to an improved measurement.

\begin{table*}\centering
    \caption{The fiducial values, the parameter ranges, and the averaged detection errors for heavy IMRIs by TianQin, LISA, and joint detection.}\label{tab:himri}
    \begin{tabular}{|m{1.6cm} || m{1.6cm} | m{1.6cm} | m{1.6cm} | m{1.6cm} | m{1.6cm} | m{1.6cm}|}
        \hline\xrowht[())]{7pt}
        Parameter & $D_L$ [Mpc] & $M$ [$\rm M_\odot$] & $q$ & $\iota$ [$\pi$] & dec & RA \\ \hline\hline\xrowht[()]{7pt}
        Fiducial & 30 & $10^6$ & 100 & 1/3 & $0^\circ$ & $12^{\rm h}$ \\ 
        \hline\xrowht[()]{7pt}
        Range & 1-1000 & $10^5$-$10^{6.2}$ & 10-5000 & 0-0.5 & $0^\circ$-$+90^\circ$ & $0^{\rm h}$-$24^{\rm h}$ \\
        \hline\xrowht[()]{7pt}
        Distr. & log & log & log & lin & lin & lin \\
        \hline\xrowht[()]{7pt}
        TianQin & $133$ & $1.5\times10^4$ & $1.1$ & $1.5\times10^{-2}$ & $5.6\times10^{-3}$ & $8.4\times10^{-3}$ \\
        \hline\xrowht[()]{7pt}
        LISA & $32$ & $5.8\times10^3$ & $3.1\times10^{-1}$ & $3.9\times10^{-3}$ & $3.7\times10^{-3}$ & $2.1\times10^{-3}$ \\
        \hline\xrowht[()]{7pt}
        Joint & $31$ & $5.3\times10^3$ & $3.0\times10^{-1}$ & $3.8\times10^{-3}$ & $1.1\times10^{-3}$ & $1.4\times10^{-3}$ \\
        \hline
    \end{tabular}
\end{table*}

\subsection{Extreme mass ratio inspirals}\label{sec:emri}

\begin{figure*}
    \centering
    \begin{minipage}{0.48\textwidth}
        \centering
        \includegraphics[width=0.95\textwidth]{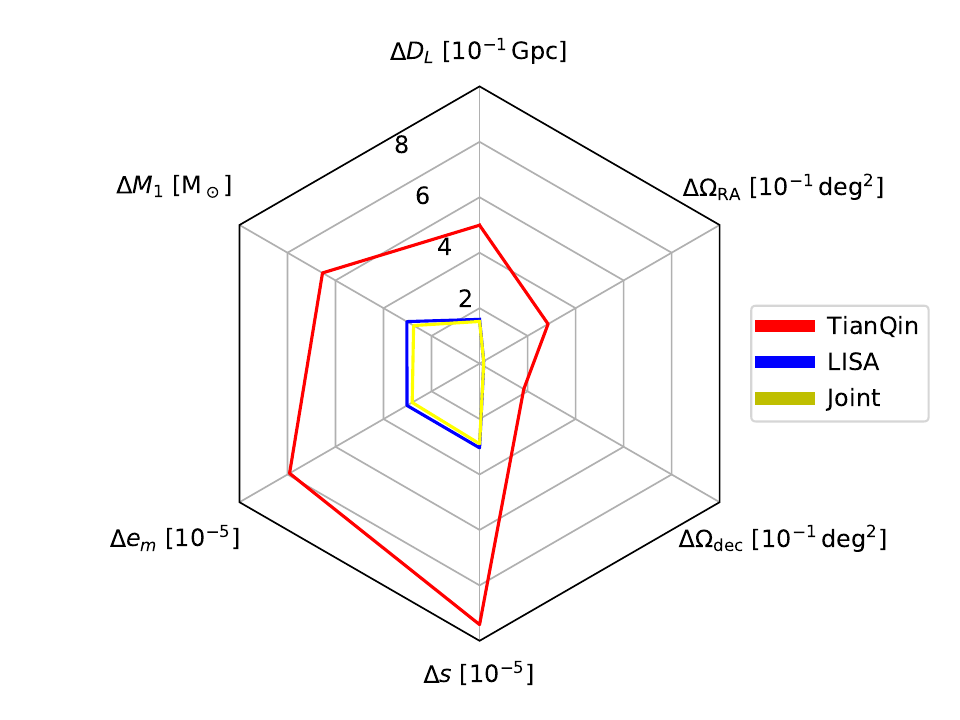}
    \end{minipage}\hfill
    \begin{minipage}{0.48\textwidth}
        \centering
        \includegraphics[width=0.95\textwidth]{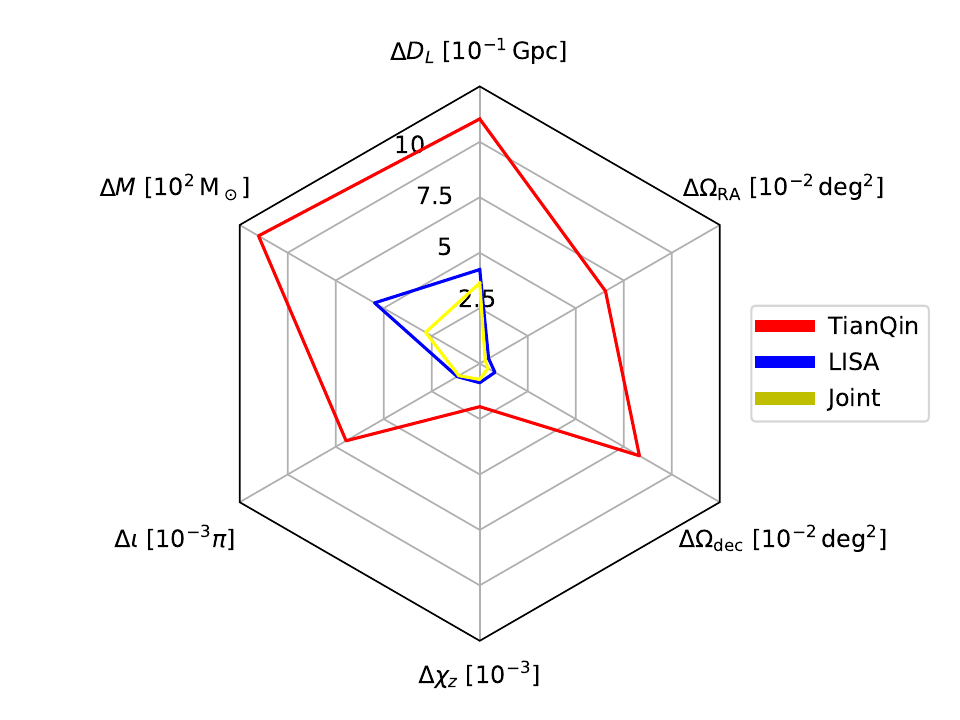}
    \end{minipage}
    \caption{
        The averaged absolute errors TianQin (red), LISA (blue), and joint detection (yellow) measure for the parameters of EMRI s(left) and MBHBs (right).
    }\label{fig:emrimbhb}
\end{figure*}

The parameters we study for an EMRI are $\Delta D_L$, the mass of the MBH $M_1$, the eccentricity at merger $e_m$, the spin of the primary BH $s$, and $\Omega$. We show their fiducial values, the parameter ranges considered and the distributions used, as well as the averaged detection errors by TianQin, LISA, and joint detection in \tref{tab:emri}. Moreover, we set the mass of the SBH orbiting the MBH to $10\,{\rm M_\odot}$. The left-hand side of \fref{fig:emrimbhb} shows that the detection accuracy of LISA is significantly better than the detection accuracy of TianQin but differs by less than one order of magnitude. Nevertheless, TianQin can contribute to slightly better constraints in the joint detection for $D_L$, $M_1$, $e_m$, and $s$. Only the accuracy of $\Omega$ in the joint detection is completely dominated by LISA which, however, can be partially attributed to the fiducial values assumed.

\begin{table*}\centering
    \caption{The fiducial values, the parameter ranges, and the averaged absolute errors of the parameter in TianQin, LISA, and joint detection for EMRIs.}\label{tab:emri}
    \begin{tabular}{|m{1.6cm} || m{1.6cm} | m{1.6cm} | m{1.6cm} | m{1.6cm} | m{1.6cm} | m{1.6cm}|}
        \hline\xrowht[())]{7pt}
        Parameter & $D_L$ [Gpc] & $M_1$ [$\rm M_\odot$] & $e_m$ & $s$ & dec & RA \\ \hline\hline\xrowht[()]{7pt}
        Fiducial & 1 & $10^6$ & 0.1 & 0.98 & $+30^\circ$ & $0^{\rm h}$ \\ 
        \hline\xrowht[()]{7pt}
        Range & 0.01-10 & $10^5$-$10^7$ & 0.002-0.5 & 0.8-0.998 & $-90^\circ$-$+90^\circ$ & $0^{\rm h}$-$24^{\rm h}$ \\
        \hline\xrowht[()]{7pt}
        Distr. & log & log & lin & lin & cos & lin \\
        \hline\xrowht[()]{7pt}
        TianQin & $5.0\times10^{-1}$ & 6.5 & $7.9\times10^{-5}$ & $9.4\times10^{-5}$ & $1.8\times10^{-1}$ & $2.9\times10^{-1}$ \\
        \hline\xrowht[()]{7pt}
        LISA & $1.6\times10^{-1}$ & 3.0 & $3.0\times10^{-5}$ & $3.0\times10^{-5}$ & $1.7\times10^{-2}$ & $1.7\times10^{-2}$ \\
        \hline\xrowht[()]{7pt}
        Joint & $1.5\times10^{-1}$ & 2.7 & $2.8\times10^{-5}$ & $2.9\times10^{-5}$ & $1.7\times10^{-2}$ & $1.7\times10^{-2}$ \\
        \hline
    \end{tabular}
\end{table*}

\subsection{Massive black hole binaries}\label{sec:mbhb}

We consider $D_L$, $M$, $\iota$, the effective spin of the BH along the angular momentum $\chi_z$, and $\Omega$ as the parameters of MBHBs. We assume the components of the binary to be of equal mass and thus the effective spin refers to any of the two MBHs. \tref{tab:mbhb} shows the fiducial values and the parameter ranges considered with the distributions applied, as well as the averaged detection error for TianQin, LISA, and joint detection. The errors are computed by varying the considered parameter over the parameter range while the other parameters are fixed to their fiducial values. On the right-hand side of \fref{fig:emrimbhb}, we see that LISA's detection error is smaller than TianQin's by a factor of two to three for most parameters. Only for $\Omega$, the difference is more significant being of more than one order of magnitude. Therefore, the contribution by TianQin to the joint detection for $\Omega$ is small but the constraints on $D_L$ and $M$ are significantly improved in joint detection compared to LISA alone. 

\begin{table*}\centering
    \caption{The fiducial values and the parameter ranges as well as TianQin's, LISA's, and joint detection's averaged absolute errors for MBHBs.}\label{tab:mbhb}
    \begin{tabular}{|m{1.6cm} || m{1.6cm} | m{1.6cm} | m{1.6cm} | m{1.6cm} | m{1.6cm} | m{1.6cm}|}
        \hline\xrowht[())]{7pt}
        Parameter & $D_L$ [Gpc] & $M$ [$\rm M_\odot$] & $\iota$ [$\pi$] & $\chi_z$ & dec & RA \\ \hline\hline\xrowht[()]{7pt}
        Fiducial & 10.5 & $4\times10^6$ & 1/3 & 0.5 & $+45^\circ$ & $1.95^{\rm h}$ \\ 
        \hline\xrowht[()]{7pt}
        Range & 0.5-50 & $10^4$-$10^7$ & 0-1 & 0.04-0.96 & $-90^\circ$-$+90^\circ$ & $0^{\rm h}$-$24^{\rm h}$ \\
        \hline\xrowht[()]{7pt}
        Distr. & log & log & cos & lin & cos & lin \\
        \hline\xrowht[()]{7pt}
        TianQin & $1.1$ & $1.2\times10^3$ & $7.0\times10^{-3}$ & $1.9\times10^{-3}$ & $8.3\times10^{-2}$ & $6.5\times10^{-2}$ \\
        \hline\xrowht[()]{7pt}
        LISA & $4.2\times10^{-1}$ & $5.5\times10^2$ & $1.2\times10^{-3}$ & $8.6\times10^{-4}$ & $7.9\times10^{-3}$ & $4.6\times10^{-3}$ \\
        \hline\xrowht[()]{7pt}
        Joint & $3.6\times10^{-1}$ & $2.8\times10^2$ & $1.1\times10^{-3}$ & $7.1\times10^{-4}$ & $4.2\times10^{-3}$ & $2.8\times10^{-3}$ \\
        \hline
    \end{tabular}
\end{table*}

\subsection{Stochastic gravitational wave background}\label{sec:sgwb}

The energy spectrum density of the SGWB can be parametrized as 
\begin{equation}
    \Omega_{\rm GW} = \Omega_0\left(\frac{f}{f_{\rm ref}}\right)^{\alpha_0} + \Omega_1\left(\frac{f}{f_{\rm ref}}\right)^{\alpha_1}\left[1 + 0.75\left(\frac{f}{f_{\rm ref}}\right)^\Delta\right]^{(\alpha_2-\alpha_1)/\Delta},
\end{equation}
where the first term represents the background formed by extra-galactic binaries~\cite{farmer_phinney_2003,regimbau_2011,moore_cole_2015} and the second term describes the foreground produced by Galactic DWDs~\cite{tq_sgwb_2022a}. $\Omega_0$ is the amplitude level of the background at the reference frequency $f_{\rm ref} = 1\,{\rm mHz}$ and $\alpha_0 = 2/3$ is the spectral index for a background of binaries. Moreover, we set $\Omega_1 = 1\times10^{-6.94}$, $\alpha_1 = 3.64$, $\alpha_2 = -3.95$, and $\Delta = 0.75$. For the dimensionless energy density of the SGWB, we consider the range $10^{-12}\leq\Omega_{\rm GW}\leq10^{-9}$ in accordance with the results of the LIGO-Virgo-KAGRA Collaboration~\cite{ligo_virgo_2021}.

\begin{figure}\centering
    \includegraphics[width=0.46\textwidth]{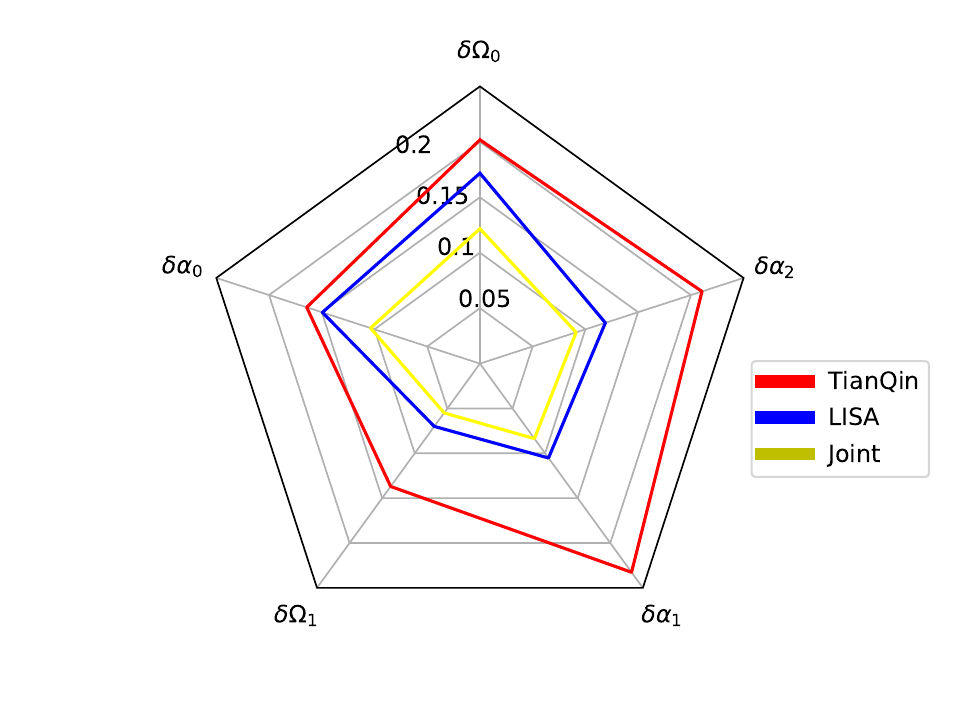}
    \caption{
        The averaged relative error for the parameters of the extra-galactic background and the galactic foreground of the SGWB measured by TianQin (red), LISA (blue), and joint detection (yellow).
    }\label{fig:sgwb}
\end{figure}

We see in \fref{fig:sgwb} that the parameters of the background $\Omega_0$ and $\alpha_0$ are constrained slightly better by LISA than by TianQin. Therefore, both contribute significantly to the joint detection of these parameters. The detection of the foreground is dominated by LISA and thus joint detection of $\Omega_1$, $\alpha_1$, and $\alpha_2$ only improves slightly by TianQin's contribution. For the detailed averaged relative errors see \tref{tab:sgwb}.

\begin{table*}\centering
    \caption{The averaged detection errors in TianQin, LISA, and joint detection for the SGWB.}\label{tab:sgwb}
    \begin{tabular}{|m{1.6cm} || m{1.6cm} | m{1.6cm} | m{1.6cm} | m{1.6cm} | m{1.6cm} | m{1.6cm}|}
        \hline\xrowht[())]{7pt}
        Parameter & $\Omega_0$ & $\alpha_0$ & $\Omega_1$ & $\alpha_1$ & $\alpha_2$ \\
        \hline\hline\xrowht[()]{7pt}
        TianQin & $2.0\times10^{-1}$ & $1.6\times10^{-1}$ & $1.4\times10^{-1}$ & $2.3\times10^{-1}$ & $2.1\times10^{-1}$ \\
        \hline\xrowht[()]{7pt}
        LISA & $1.7\times10^{-1}$ & $1.5\times10^{-1}$ & $7.0\times10^{-2}$ & $1.1\times10^{-1}$ & $1.2\times10^{-1}$ \\
        \hline\xrowht[()]{7pt}
        Joint & $1.2\times10^{-1}$ & $1.0\times10^{-1}$ & $5.5\times10^{-2}$ & $8.4\times10^{-2}$ & $9.1\times10^{-2}$ \\
        \hline
    \end{tabular}
\end{table*}

\section{Summary}\label{sec:sum}

We study how accurately the parameters of DWDs, SBHBs, light and heavy IMRIS, EMRIs, MBHBs, and the SGWB formed by unresolved binaries are detected by TianQin, LISA, and joint detection. For each source, we consider a range of realistic parameters and compute the averaged detection error. We find that TianQin's detection accuracy is particularly good for DWDs and light IMRIs while LISA obtains tighter constraints for heavy IMRIs, EMRIs, MBHBs, and the galactic foreground of the SGWB. For SBHBs and the extra-galactic SGWB, TianQin and LISA obtain comparable results thus contributing similarly to joint detection. However, for all sources, joint detection yields improved detection accuracies showing the importance of coordinated detections. This study focuses on the averaged detection accuracy but detection errors can vary significantly depending on the specific value of the parameter. Therefore, we refer interested readers to Ref.~\cite{torres-orjuela_huang_2023} where the detection accuracy for TianQin, LISA, and joint detection are discussed in more detail. Although the discussion in Ref.~\cite{torres-orjuela_huang_2023} is much more extensive, the topic of joint detection requires further studies to understand the potential profit of coordinated detections by TianQin and LISA.

%
%
\bibliographystyle{unsrt}
\bibliography{ref}
\end{document}